\documentclass[aps,prd,preprint,tightenlines,superscriptaddress,nofootinbib]{revtex4}
\usepackage{epsfig}
\usepackage{graphicx}
\usepackage{psfrag}
\usepackage{amsmath,amssymb}
\usepackage{colordvi}
\usepackage{amsfonts}
\usepackage{enumerate}
\usepackage{slashed}
\usepackage{color}
\usepackage{xcolor}

\begin{document}

\preprint{MIT-CTP 4962}

\title{Transverse Momentum Dependent Quasi-Parton-Distributions}

\author{Xiangdong Ji}
\affiliation{Tsung-Dao Lee Institute, and College of Physics and Astronomy,
Shanghai Jiao Tong University, Shanghai, 200240, China}
\affiliation{Maryland Center for Fundamental Physics,
Department of Physics, University of Maryland,
College Park, Maryland 20742, USA}

\author{Lu-Chang Jin}
\affiliation{Department of Physics, University of Connecticut, Storrs, CT 06269, USA}
\affiliation{RIKEN-BNL Research Center, Brookhaven National Laboratory, Building 510, Upton, NY 11973}

\author{Feng Yuan}
\affiliation{Nuclear Science Division, Lawrence Berkeley
National Laboratory, Berkeley, CA 94720, USA}

\author{Jian-Hui Zhang}
\affiliation{Institut f\"ur Theoretische Physik, Universit\"at Regensburg, D-93040 Regensburg, Germany}

\author{Yong Zhao}
\affiliation{Center for Theoretical Physics, Massachusetts Institute of Technology, Cambridge, MA 02139, USA}

\date{\today}
\vspace{0.5in}
\begin{abstract}
We investigate the transverse momentum dependent parton distributions (TMDs)
in the quasi-parton-distribution framework. The long-standing hurdle of the
so-called pinch pole singularity from the space-like gauge links in the TMD
definitions can be resolved by the finite length of the gauge link along the 
hadron moving direction. In addition, with the soft factor subtraction, the 
quasi-TMD is free of linear divergence. 
We further demonstrate that the energy evolution equation of the quasi-TMD
{{a.k.a.}} the Collins-Soper evolution, only depends on the hadron momentum. 
This leads to a clear matching between the quasi-TMD and the standard TMDs.
\end{abstract}

\maketitle

\section{Introduction}
Transverse momentum dependent parton distributions (TMDs) are one
of the major focuses in nucleon tomography studies at existing and
future facilities~\cite{Boer:2011fh}. Theoretically, they have attracted
great interest starting in early 80's, and considerable developments have been achieved
in recent years~\cite{Collins,Collins:1981uk,Ji:2004wu,bbdm}.
Pioneering work to 
compute the TMD matrix elements from lattice QCD has also been performed in 
Ref.~\cite{Musch:2010ka}, where the 
longitudinal momentum fraction $x$ for the quarks 
has been integrated out. Such results have generated
interest in computing TMDs from lattice QCD in hadron physics community. 

In the last few years, there has been great progress on computing parton physics from lattice QCD, 
thanks to the large momentum effective theory (LaMET)~\cite{Ji:2013dva}. LaMET is based on 
the observation that parton physics defined in terms of lightcone correlations can be obtained from 
time-independent Euclidean correlations, now known as quasi-distributions, boosted to the infinite 
momentum frame. For a finite but large momentum feasible on the lattice, the two quantities are not 
identical, but they can be connected to each other by a perturbative matching relation, up to power 
corrections that are suppressed by the hadron momentum. LaMET has been applied to computing various 
PDFs~\cite{Lin:2014zya,Alexandrou:2015rja,Chen:2016utp,Alexandrou:2016jqi,Chen:2017mzz,
Lin:2017ani,Alexandrou:2017dzj} as well as meson DAs~\cite{Zhang:2017bzy,Chen:2017gck} 
(see also~\cite{Ma:2014jla,Ma:2017pxb} for slightly different proposals). In addition, 
theoretical developments have been achieved on the renormalization of 
the quasi-parton-distribution-functions (Q-PDFs) and on their matching to the usual PDFs~\cite{Xiong:2013bka,Xiong:2017jtn,Wang:2017qyg,Wang:2017eel,Stewart:2017tvs,Ji:2015qla,
Xiong:2015nua,Ji:2015jwa,Ishikawa:2016znu,Chen:2016fxx,
Constantinou:2017sej,Alexandrou:2017huk,Chen:2017mzz,Ji:2017oey,Ji:2017rah,
Ishikawa:2017faj,Green:2017xeu,Li:2016amo,Monahan:2016bvm,Briceno:2017cpo,Monahan:2017hpu,Zhang:2018ggy,Izubuchi:2018srq}. 
Unfortunately, there has been no lattice effort to compute the TMDs from the quasi-TMDs (Q-TMDs). The major
hurdle is that the formulation of the TMDs is different from the integrated 
PDFs and, in particular, the gauge links associated with the Q-TMDs lead
to the so-called pinch pole singularities. This is a generic feature of the
TMDs defined with a space-like gauge link~\cite{bbdm,Collins:1981uk}.
We have to either subtract or regulate
these singularities before we can make meaningful computations of 
the Q-TMDs on the lattice~\cite{Collins}.
In Ref.~\cite{Ji:2014hxa}, a soft factor subtraction involving transverse
gauge links has been proposed to formulate the Q-TMDs. However,
this formalism may have practical difficulties for lattice computations at present. 

In this paper, we will reinvestigate the TMDs in LaMET or Q-TMDs framework. 
We will show that, with finite length gauge links in the Q-TMDs, there will be no 
pinch pole singularity. This will pave
the way to perform the TMD calculations on the lattice. Moreover,
with an explicit one-loop calculation, we demonstrate that the
energy evolution of the TMDs depends on the hadron momentum.
This will clarify an important issue to match the Q-TMDs
to the standard TMDs extracted from the experiments. 

\begin{figure}[tbp]
\begin{center}
\includegraphics[width=11cm]{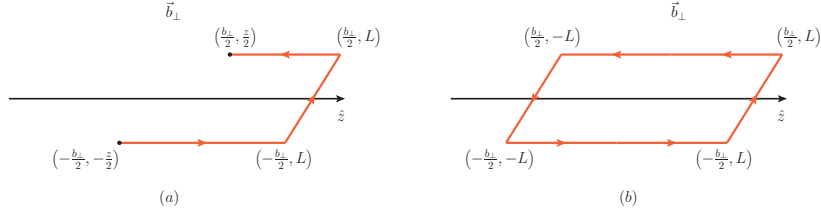}
\end{center}
\caption[*]{Illustration of the gauge links in the un-subtracted Quasi-TMD (a)
and the soft factor (b).}
\label{gaugelink}
\end{figure}

Our focus will be on the basics of the formalism and setting up the foundation for future 
numerical simulations on the lattice. Let us start with the un-subtracted Q-TMD quark distribution 
defined with finite length gauge links, 
\begin{eqnarray}
q(x_z,{\vec k}_\perp;L)|^{(unsub.)}&=&\frac{1}{2}\int\frac{dz\, d^2{\vec b}_\perp}{(2\pi)^3}e^{-ik_z z-i\vec{k}_\perp\cdot \vec{b}_\perp}\langle PS|\overline{\psi}(-\frac{{\vec b}_\perp}{2},-\frac{z}{2})
{\cal L}_{n_z(-\frac{{\vec b}_\perp}{2},-\frac{z}{2};-\frac{{\vec b}_\perp}{2},\pm L)}^\dagger\gamma^z \nonumber\\
&&\times  {\cal L}_{T(-\frac{{\vec b}_\perp}{2},\pm L;\frac{{\vec b}_\perp}{2},\pm L)}^\dagger
{\cal L}_{n_z(\frac{{\vec b}_\perp}{2},\frac{z}{2};\frac{{\vec b}_\perp}{2},\pm L)}\psi(\frac{{\vec b}_\perp}{2},\frac{z}{2})|PS\rangle \ , \label{tmdq}
\end{eqnarray}
where $(\vec{b}_\perp,z)$ represents the 3-dimensional coordinate space variable separated by
the quark and antiquark fields, $x_z=k_z/P_z$ and the proton is moving along $+\hat z$ direction,
${\vec k}_\perp$ represents the transverse momentum of the quark. 
In the above definition, ${\cal L}_{n_z({\vec y}_\perp, z_1;{\vec y}_\perp, z_2)}={\cal P}\,exp\left[-ig\int_{z_2}^{z_1}
d\lambda\, n_z\cdot A(\lambda n_z+{\vec y}_\perp)\right]$ represents the gauge link along the $\hat z$ direction 
with the large length $L\gg |z|$, where the 4-vector $n_z$ is defined as $n_z^\mu=(0,0,0,1)$.
We have also included a transverse gauge link to make
the gauge links connected as shown in Fig.~\ref{gaugelink}(a). 
 
In the TMD formalism, it has been demonstrated that the soft factor subtraction plays
an important role to properly address the relevant factorization properties~\cite{Collins}. 
In this paper, we introduce the following soft factor subtraction, 
\begin{equation}
q^{(sub.)}(x_z,{\vec b}_\perp)=\frac{q^{(unsub.)}(x_z,{\vec b}_\perp;L)}{\sqrt{S^{n_z, n_z}({\vec b}_\perp;L)}}\ , \label{tmd}
\end{equation}
where $q^{(unsub.)}(x_z,{\vec b}_\perp; L)$ is the un-subtracted Q-TMD
in Eq.~(\ref{tmdq}) in the Fourier transform ${\vec b}_\perp$-space with respect to the
transverse momentum ${\vec k}_\perp$, and $S^{n_z, n_z}({\vec b}_\perp; L)$ is defined as
\begin{eqnarray}
S^{n_z, n_z}({\vec b}_\perp;L)&=&{\langle 0|{\cal L}_{T(\frac{{\vec b}_\perp}{2},-L;-\frac{{\vec b}_\perp}{2},-L)}^\dagger{\cal L}_{n_z(\frac{{\vec b}_\perp}{2},0;\frac{{\vec b}_\perp}{2},-L)}^\dagger
{\cal L}_{n_z(\frac{{\vec b}_\perp}{2},L;\frac{{\vec b}_\perp}{2},0)}^\dagger}\nonumber\\
&&\times {{\cal L}_{T(\frac{{\vec b}_\perp}{2},L;-\frac{{\vec b}_\perp}{2},L)} {\cal
L}_{ n_z(-\frac{{\vec b}_\perp}{2},L;-\frac{{\vec b}_\perp}{2},0)} {\cal L}_{n_z(-\frac{{\vec b}_\perp}{2},0;-\frac{{\vec b}_\perp}{2},-L)} |0\rangle   }\, , \label{softg}
\end{eqnarray}
with ${\cal L}_{n_z}$ being the longitudinal gauge link along the $\hat z$ direction and ${\cal L}_T$
the transverse gauge link at $z=L$ and $z=-L$, as shown in Fig.~\ref{gaugelink} (b). In other words, the above
soft factor is just a Wilson loop. 

The rest of this paper is organized as follows. In Sec.~II, we will
show the absence of pinch pole singularity in the Q-TMD with finite length gauge links with an explicit calculation
at one-loop order. In Sec.~III, we will discuss the matching between
the Q-TMD and the standard TMD. We then summarize our paper in Sec. IV.

\section{Absence of the Pinch Singularity in Q-TMDs}

\begin{figure}[tbp]
\begin{center}
\includegraphics[width=10cm]{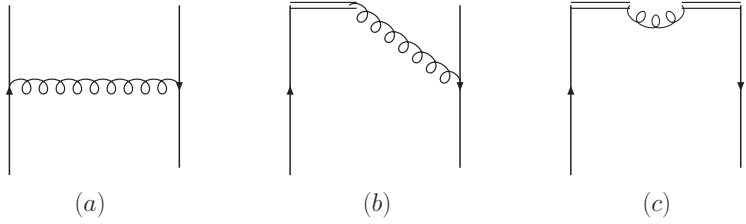}
\end{center}
\caption[*]{Real diagram contributions to the Q-TMD quark distributions 
at one-loop order. The complex conjugate of Diagram (b) is implied. Diagram (c) 
would contain the pinch pole singularity with infinite length gauge links in the
TMD definition. However, with a finite length $L$, this singularity is absent in Q-TMD.}
\label{pdfr}
\end{figure}

\begin{figure}[tbp]
\begin{center}
\includegraphics[width=10cm]{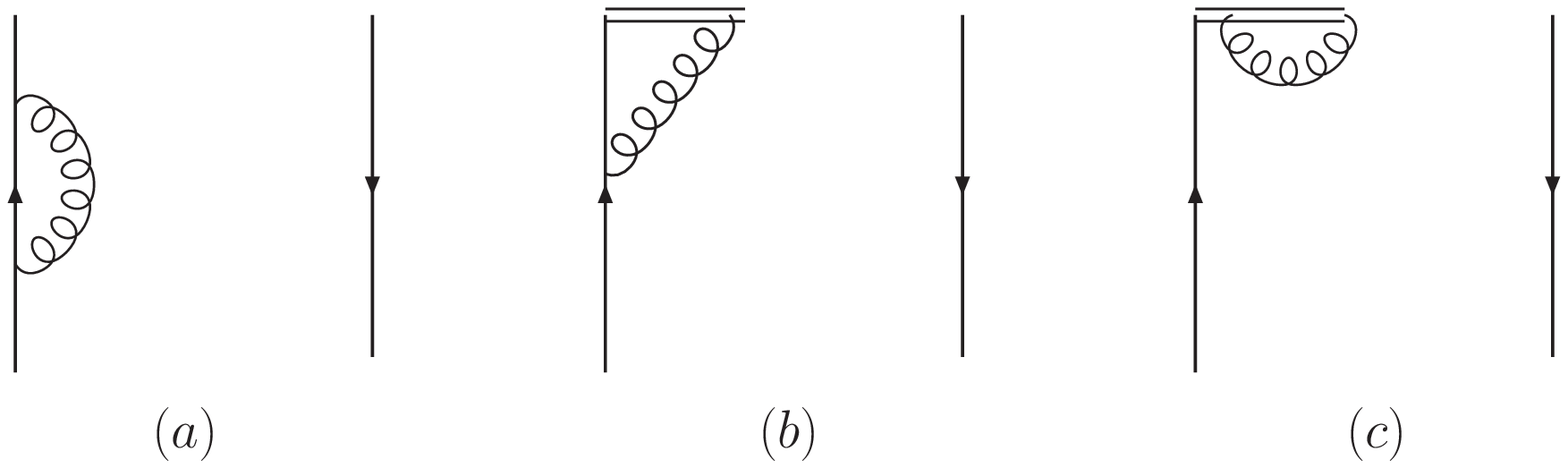}
\end{center}
\caption[*]{Virtual diagram contributions to the Q-TMD quark distributions at one-loop order. 
The complex conjugate is also implied. Diagram (c) requires the renormalization of the 
gauge link self-interaction. This follows recent examples in the
collinear Q-PDF cases.}
\label{pdfv}
\end{figure}

To show that we do not encounter the pinch pole singularity, we will 
carry out a one-loop calculation. We take the example of 
quark Q-TMD on a quark target. In Feynman gauge, the one-loop diagrams are
shown in Figs.~\ref{pdfr} and \ref{pdfv}. The final result can also serve
as a matching between the Q-TMD and the standard TMD.
Because of the finite length of the gauge links, the eikonal propagator
in these diagrams will be modified accordingly,
\begin{equation}
(-ig)\frac{i n^\mu }{n\cdot k\pm i\epsilon}\Longrightarrow
(-ig)\frac{i n^\mu}{n\cdot k}\left(1-e^{\pm in\cdot k L}\right) \ ,\label{eikonal}
\end{equation}
where $n^\mu$ represents the gauge link direction. In the present case $n^\mu=n_z^\mu$.
In perturbative calculations, we will make use of the large
length limit $|LP_z|\gg 1$. By doing that, many of previous results can
be applied to our calculations. 
For example, in the large $L$ limit, we have the
following identity: 
$\lim_{L\to \infty}\frac{1}{n\cdot k}e^{\pm i Ln\cdot k}=\pm i\pi \delta(n\cdot k) $.

At one-loop order, the pinch pole singularity could potentially come from the diagram $(c)$
of Fig.~\ref{pdfr} in the limit of infinite gauge link with $L\to \infty$, 
\begin{align}
&q^{(1)}(x_z,{\vec k}_\perp)|_{\ref{pdfr}(c)}^{L\to \infty}\nonumber\\
&=\frac 1 2\int\frac{dk^0 dk_z}{(2\pi)^4}\bar u(p)\gamma^z (ig t^a)(ig t^a) \frac{-i}{n\cdot(P-k)-i\epsilon}\frac{i}{n\cdot(P-k)+i\epsilon}\frac{-i}{(P-k)^2} u(p)\delta\big(k_z-x_z P_z\big)\nonumber\\
&=\frac{\alpha_s}{4\pi^2}C_F
\frac{P_z}{\sqrt{(1-x_z)^2P_z^2+{\vec k}_\perp^2}}\frac{1}{(1-x_z)P_z+i\epsilon}
\frac{1}{(1-x_z)P_z-i\epsilon}\ .
\end{align}
For the case of integrated parton distributions, we integrate over ${\vec k}_\perp$ to 
obtain the one-loop result. 
However, in the current case, we have to keep the transverse momentum ${\vec k}_\perp$.
In addition, we note that the above contribution is power suppressed by ${\vec k}_\perp/P_z$
for $x_z\neq1$. That means to leading power in $P_z$ this diagram only contributes to $\delta(1-x_z)$,
and can be written as
\begin{eqnarray}
q^{(1)}(x_z,{\vec k}_\perp)|_{\ref{pdfr}(c)}^{L\to \infty}&=&\frac{\alpha_s}{4\pi^2}C_F
\delta(1-x_z)\int\frac{dk_z}{\sqrt{k_z^2+{\vec k}_\perp^2}}\frac{1}{k_z+i\epsilon}\frac{1}{k_z-i\epsilon} \ .
\end{eqnarray}
The above integral is not well-defined, because the two poles
are pinched. We are forced to take the pole at $k_z=0$, which is, however, 
divergent. This is a common issue for parton distributions
defined with gauge links along the space-like direction~\cite{bbdm,Collins:1981uk}.

With finite length gauge links, the above result will be modified to
\begin{eqnarray}\label{fig2cfiniteL}
q^{(1)}(x_z,{\vec k}_\perp)|_{\ref{pdfr}(c)}&=&\frac{\alpha_s}{4\pi^2}C_F
\frac{P_z}{\sqrt{(1-x_z)^2P_z^2+{\vec k}_\perp^2}}\frac{1}{(1-x_z)P_z}\frac{1}{(1-x_z)P_z}\nonumber\\
&&\times \left(1-e^{ i(1-x_z)P_zL}\right)\left(1-e^{ -i(1-x_z)P_zL}\right)\ ,
\end{eqnarray}
where we have used Eq.~(\ref{eikonal}). We find again that this result is power 
suppressed for $x_z\neq 1$. Therefore, we can simplify the above
equation as
\begin{eqnarray}
q^{(1)}(x_z,{\vec k}_\perp)|_{\ref{pdfr}(c)}&=&\frac{\alpha_s}{4\pi^2}C_F\delta(1-x_z)\int\frac{dk_z}{k_z^2}\frac{1}{\sqrt{k_z^2+{\vec k}_\perp^2}}
\left(1-e^{ ik_zL}\right)\left(1-e^{ -ik_zL}\right)\ .\label{e13}
\end{eqnarray}
Now, the integral is well regulated around $k_z=0$. Moreover, it does not contribute to the infrared behavior of the 
Q-TMD at low transverse momentum, as can be seen by an explicit integration over small ${\vec k}_\perp$ in Eq.~(\ref{e13}), 
which does not yield any divergence. 
In particular, taking the Fourier transform with respect to ${\vec k}_\perp$, we obtain the
following expression in the ${\vec b}_\perp$-space,
\begin{eqnarray}
q^{(1)}(x_z,{\vec b}_\perp)|_{\ref{pdfr}(c)}&=&\frac{\alpha_s}{2\pi}C_F\delta(1-x_z) 2{\cal K}(\xi_b) \ ,
\end{eqnarray}
where $\xi_b=L/|{\vec b}_\perp|$ and the function ${\cal K}$ is defined as
\begin{equation}
{\cal K}(\xi_b)=2\xi_b \tan^{-1}\xi_b-\ln(1+\xi_b^2) \ .
\end{equation}
At large $\xi_b$ the above equation goes like $\pi\xi_b-2\ln\xi_b$, while
at small $\xi_b$ it behaves as $\xi_b^2$.

Furthermore, with the soft factor subtraction, we will be able to eliminate
the $1/|{\vec k}_\perp|$ term at small ${\vec k}_\perp$ in Eq.~(\ref{e13}). The subtraction term relevant to Fig.~\ref{pdfr}(c) comes from the gluon exchange between the two longitudinal Wilson lines with length $2L$ in Fig.~\ref{gaugelink}(b). It can be computed in the same way as that of Fig.~\ref{pdfr}(c) and leads to the same result as Eq.~(\ref{fig2cfiniteL}) except that $L$ needs to be replaced by $2L$. We then have the following result after subtraction,
\begin{eqnarray}
q^{(1)}(x_z,{\vec k}_\perp)|_{\ref{pdfr}(c)}^{(sub.)}&=&q^{(1)}(x_z,{\vec k}_\perp)|_{\ref{pdfr}(c)}^{(sub.)}-\frac{1}{2}q^{(1)}(x_z,{\vec k}_\perp)|_{\ref{pdfr}(c)}^{(sub.)}(L\to 2L)\nonumber\\
&=&\frac{\alpha_s}{4\pi^2}
C_F\delta(1-x_z)\int\frac{dk_z}{k_z^2}\frac{1}{\sqrt{k_z^2+{\vec k}_\perp^2}}
\frac{1}{2}\left[\left(1-e^{ ik_zL}\right)^2+\left(1-e^{ -ik_zL}\right)^2\right]\  .\label{e13p}
\end{eqnarray}
In the Fourier transform ${\vec b}_\perp$-space, the above result becomes,
\begin{eqnarray}
q^{(1)}(x_z,{\vec b}_\perp)|_{\ref{pdfr}(c)}^{(sub.)}&=&\frac{\alpha_s}{2\pi}C_F\delta(1-x_z) \left[2{\cal K}(\xi_b)-{\cal K}(2\xi_b)\right] \ ,
\end{eqnarray}
where the second term comes from the soft factor. Clearly, the linear term of $\xi_b$ is cancelled out in the subtracted contribution, and the result goes like $\ln(\xi_b^2)$ at large $\xi_b$, whereas at small $\xi_b$ it again behaves like $\xi_b^2$.

Similarly, the contribution
from Fig.~\ref{pdfv}(c) is given by (for a finite length gauge link)
\begin{align}
q^{(1)}(x_z,{\vec k}_\perp)|_{\ref{pdfv}(c)}&=\frac{-ig^2 C_F}{2}\int\frac{d^4k'}{(2\pi)^4}\frac{1}{(p-k')^2}\frac{1}{[n\cdot(p-k')]^2}\nonumber\\
&\times\left(1-e^{ in\cdot(p-k')L}\right)\left(1-e^{ -in\cdot(p-k')L}\right)\delta(k'_z-x_z P_z)\delta^{(2)}({\vec k}_\perp).
\end{align}
After soft factor subtraction, it gives the following expression,
\begin{eqnarray}
q^{(1)}(x_z,{\vec k}_\perp)|_{\ref{pdfv}(c)}^{(sub.)}&=&\frac{\alpha_s}{8\pi^2}
C_F\delta(1-x_z)\delta^{(2)}({\vec k}_\perp)\int\frac{dk_zd^2{\vec k}_\perp'}{k_z^2}\left(\frac{1}{\sqrt{{\vec k}_\perp^{\prime 2}}}-\frac{1}{\sqrt{k_z^2+{\vec k}_\perp^{\prime 2}}}\right)\nonumber\\
&&\times \left[\left(1-e^{ ik_zL}\right)\left(1-e^{ -ik_zL}\right)-\frac{1}{2}\left(1-e^{ i2k_zL}\right)\left(1-e^{ -i2k_zL}\right)\right] \ ,\label{e13p2}
\end{eqnarray}
where we have rewritten the integral over ${\vec k}_\perp'$ in such a way that the linear divergence is
manifestly absent. 
In addition, all $L$-dependent contributions that are not suppressed in the large $L$ limit cancel out in the full subtracted Q-TMD. 
The cancellation occurs either among the unsubtracted Q-TMD diagrams or with similar contributions from the soft factor. This 
can be easily seen from computations in coordinate space. 

As there is no linear divergence associated with the gauge links after soft factor subtraction, 
we can work in dimensional regularization, which leads to the following contribution,
\begin{eqnarray}\label{3csub}
q^{(1)}(x_z,{\vec b}_\perp)|_{\ref{pdfv}(c)}^{(sub.)}&=&\frac{\alpha_s}{4\pi}
C_F\delta(1-x_z)\big[\ln\frac{L^2\mu^2}{4c_0^2}+2\big] \ ,
\end{eqnarray}
in the Fourier transform ${\vec b}_\perp$-space, where the UV divergence
has been subtracted with $\overline{\rm MS}$ scheme. In the dimensional
regulation, the linear divergence is not manifest explicitly in the unsubtracted Q-TMD, 
and the result is the same as above with a factor of 2.
However, if a cutoff scheme is chosen, there will be an explicit linear divergence
in the unsubtracted Q-TMD,  
\begin{eqnarray}
q^{(1)}(x_z,{\vec b}_\perp)|_{\ref{pdfv}(c), tot.}^{(unsub.)}&=&\frac{\alpha_s}{2\pi}
C_F\delta(1-x_z)\left[4-\frac{2\pi L}{a}+2\ln\frac{L^2}{a^2}\right] \ ,
\end{eqnarray}
whereas the linear divergence is cancelled out for the subtracted Q-TMD 
\begin{eqnarray}
q^{(1)}(x_z,{\vec b}_\perp)|_{\ref{pdfv}(c), tot.}^{(sub.)}&=&\frac{\alpha_s}{2\pi}
C_F\delta(1-x_z)\left[2+\ln\frac{L^2}{4a^2}\right] \ .\label{e16}
\end{eqnarray}
The transverse gauge link contribution can be calculated 
in complete analogy and we have 
\begin{eqnarray}
q^{(1)}(x_z,{\vec b}_\perp)|_{\ref{pdfv}(c)}^{(unsub.)T}&=&\frac{\alpha_s}{2\pi}
C_F\delta(1-x_z)\left[2-\frac{\pi {\vec b}_\perp}{a}+ \ln\frac{{\vec b}_\perp^2}{a^2}\right] \ .
\end{eqnarray}
For the subtracted Q-TMD, the transverse gauge link contribution is cancelled out
completely,  
\begin{eqnarray}\label{3csubT}
q^{(1)}(x_z,{\vec b}_\perp)|_{\ref{pdfv}(c)}^{(sub.)T}&=&0 \ .
\end{eqnarray}

Eqs.~(\ref{3csub}) to (\ref{e16}) are independent of ${\vec b}_\perp$, and therefore will remain the same at large or small $\xi_b$. From the results above, one can easily see that the 
subtracted result of Fig.~\ref{pdfr}(c),~\ref{pdfv}(c) has a residual logarithmic UV divergence, 
\begin{align}
q^{(1)}(x_z,{\vec b}_\perp)|_{\ref{pdfr}(c),\ref{pdfv}(c)}^{(sub.)}&=\frac{\alpha_s}{2\pi}
C_F\delta(1-x_z)\big[\ln\frac{L^2\mu^2}{4c_0^2}+2+2{\cal K}(\xi_b)-{\cal K}(2\xi_b)\big] \ ,\label{e13p3new}
\end{align}
in the Fourier transform ${\vec b}_\perp$-space with respect to the transverse momentum 
${\vec k}_\perp$, where $c_0=2e^{-\gamma_E}$. In the above equation, 
we have applied the dimensional regulation for the UV divergence
and renormalize in the $\overline{\rm MS}$ scheme with scale $\mu$. If a lattice regulator is adopted, we will
obtain the same expression with $\mu/c_0=1/a$, where $a$ is the lattice 
spacing parameter. Because of the above contribution, we will have an additional anomalous
dimension contribution from Eq.~(\ref{e13p3new}) for the evolution 
equation of the Q-TMD. 

The rest of the real diagrams in Fig.~\ref{pdfr} can be calculated by 
safely taking the large $L$ limit. For example, the contribution of 
Fig.~\ref{pdfr}(b) is given by
\begin{align}
\frac {-i g^2 C_F}{2}\int\frac{dk^0 dk_z}{(2\pi)^4}\bar u(p)\gamma^z \frac{1}{n\cdot(P-k)}\frac{1}{\slashed k}\gamma^z\frac{1}{(P-k)^2} u(p)
\left(1-e^{ in\cdot(P-k)L}\right)\delta\big(k_z-x_z P_z\big)\ ,
\end{align}
and leads to the following result
\begin{align}
q^{(1)}(x_z,{\vec k}_\perp)|_{\rm Fig.~2(b)}=&\frac{\alpha_s}{4\pi^2}C_F\left[\frac{1}{{\vec k}_\perp^2}\frac{x_z}{1-x_z}\left(\frac{P_z(1-x_z)}{\sqrt{{\vec k}_\perp^2+P_z^2(1-x_z)^2}} + \frac{P_zx_z}{\sqrt{{\vec k}_\perp^2+P_z^2x_z^2}}\right)\right.\nonumber\\
&\left.+{1\over 1-x_z}{1\over P_z^2}\left({P_z\over \sqrt{{\vec k}_\perp^2 + P_z^2x_z^2}}-{P_z\over \sqrt{{\vec k}_\perp^2 + P_z^2(1-x_z)^2}}\right)\right]\left(1-e^{ i(1-x_z)P_zL}\right) \ . \label{e6}
\end{align}
However, this additional factor $e^{[i(1-x_z)P_zL]}$ does not
contribute in the large $L$ limit.  
It is interesting to note that if we take $P_z\to \infty$ first, the above
equation will lead to a divergence of $1/(1-x_z)$, which is same as the 
light-cone singularity in the usual TMD definition. 
Again, the contributions from the regions of $x_z<0$ and $x_z>1$ are power suppressed in the limit
 $|{\vec k}_\perp|\ll P_z$. The final result from this diagram can be written as, 
\begin{equation}
\frac{\alpha_s}{2\pi^2}C_F\frac{1}{{\vec k}_\perp^2}\left[\frac{2x_z}{(1-x_z)_+}\theta(x_z)\theta(1-x_z)+\delta(1-x_z)\ln\frac{\zeta^2}{{\vec k}_\perp^2}\right] \ ,\label{e7}
\end{equation}
where $\zeta^2=x_z^2(2n_z\cdot P)^2/(-n_z^2)=4x_z^2P_z^2$ and we have applied a 
principal-value prescription to evaluate the second term in Eq.~(\ref{e7}).

Because there is no gauge link contribution from Fig.~\ref{pdfr} (a), its result 
will be the same as previously calculated in Ref.~\cite{Ji:2014hxa}
\begin{align}
q^{(1)}(x_z,{\vec k}_\perp)|_{\rm Fig.~2(a)}&=\frac{\alpha_s}{4\pi^2}C_F\frac{1-\epsilon}{{\vec k}_\perp^2}\frac{(1-x_z)\big(\sqrt{{\vec k}_\perp^2+P_z^2(1-x_z)^2}+P_z(1-x_z)\big)}{\sqrt{{\vec k}_\perp^2+P_z^2(1-x_z)^2}}.
\end{align}
 In the limit $|{\vec k}_\perp|\ll P_z$, the above result reduces to
\begin{equation}\label{e2ared}
\frac{\alpha_s}{2\pi^2}C_F\frac{1-\epsilon}{{\vec k}_\perp^2}(1-x_z)\ .
\end{equation}

Similar calculations can be
performed for the virtual diagrams of Fig.~\ref{pdfv}(a,b), and the result reads~\cite{Ji:2014hxa}
\begin{align}\label{e3ab}
q^{(1)}(x_z,{\vec b}_\perp)|_{3(a), 3(b)}&=\frac{\alpha_s}{2\pi}C_F\delta(1-x_z)\Big[-\frac{1}{\epsilon^2}-\frac{3}{2\epsilon}+\frac{1}{\epsilon}\ln\frac{\zeta^2}{\mu^2}+\ln\frac{\zeta^2}{\mu^2}-\frac{1}{2}\Big(\ln\frac{\zeta^2}{\mu^2}\Big)^2+\frac{\pi^2}{12}-2\Big].
\end{align}

Finally, the total contribution of the subtracted
Q-TMD quark distribution at one-loop order can be obtained from Eqs.~(\ref{e13p3new}), (\ref{e3ab}) and the Fourier transform of (\ref{e7}), (\ref{e2ared}),
\begin{eqnarray}
{q}_{QTMD}^{(sub.)(1)}(x_z,{\vec b}_\perp;\zeta^2)&=&\frac{\alpha_s}{2\pi}C_F\left\{\left(-\frac{1}{\epsilon}
+\ln\frac{c_0^2}{{\vec b}_\perp^2\mu^2}\right){\cal P}_{q/q}(x_z)+(1-x_z)\right.\nonumber\\
&&\left.+\delta(1-x_z)\left[\frac{3}{2}\ln\frac{{\vec b}_\perp^2\mu^2}{c_0^2}+\ln\frac{\zeta^2L^2}{4c_0^2}
-\frac{1}{2}\left(\ln\frac{\zeta^2{\vec b}_\perp^2}{c_0^2}\right)^2\right.\right.\nonumber\\
&&\left.\left.+2{\cal K}(\xi_b)-{\cal K}(2\xi_b)\right]\right\}  \label{oneloop}
\end{eqnarray}
in ${\vec b}_\perp$-space, where
$\mu$ is the renormalization scale in the $\overline{\rm MS}$ scheme,
and ${\cal P}_{q/q}(x_z)=\left(\frac{1+x_z^2}{1-x_z}\right)_+$ is the usual splitting kernel for the quark. 
We would like to emphasize a number of important points here. First, the Q-TMDs
only have contributions in the region $0<x_z<1$. This is because, as mentioned
above, we are taking the physical limit for TMD, i.e., $P_z\gg |{\vec k}_\perp|$. In this limit, the
contributions in the region $x_z>1$ and $x_z<0$ are power suppressed.
Second, similar to the previous formalisms for the TMDs, the Q-TMDs contain
the double logarithms as indicated in the above equation. From the explicit
calculations, we find that these double logarithms depend on the hadron
momentum $P_z$ in the Q-PDF framework. Therefore, the 
associated energy evolution, i.e., the Collins-Soper
evolution, will depend on $P_z$ not $L$. Finally, as expected,
the Q-TMD at one-loop order contains infrared divergence,
which corresponds to the collinear splitting of the quark.  

Comparing to the result in Ref.~\cite{Ji:2014hxa}, we find an additional term 
from the soft factor subtraction in the Q-TMD. This term will lead to 
a different matching between the Q-TMD and the standard TMD.

\section{Matching to the Standard TMDs}

With the above one-loop result for the Q-TMD quark distribution, we can
match to the usual TMDs at this order following the procedure of Ref.~\cite{Ji:2013dva}.
However, there is scheme dependence in the usual TMDs to regulate
the relevant light-cone singularities~\cite{Collins}. Therefore, a direct matching 
to the various TMDs will introduce the scheme dependence as well.
On the other hand, as demonstrated in 
Refs.~\cite{Catani:2000vq,Catani:2013tia,Prokudin:2015ysa}, all TMD schemes
lead to the same result after resumming the large logarithms. Therefore,
it is more appropriate to carry out the matching between the Q-TMDs
and the standard TMDs after the resummation has been performed.

This resummation is carried out by solving the associated evolution
equations~\cite{Collins}. For the Q-TMD quark distribution, the relevant 
Collins-Soper evolution can be derived~\cite{Ji:2014hxa}, and the complete resummation 
result can be expressed in terms of the integrated parton distributions~\cite{Collins},
\begin{eqnarray}\label{tmdaspdf}
{q}_{QTMD}(x_z,{\vec b}_\perp;\zeta^2)&=& e^{-{ {S}^q(\zeta,{\vec b}_\perp)}}e^{-{ {S}_w^q(\zeta,\mu_L)}}\int\frac{dx'}{x'}f_q(x',\mu_b)\nonumber\\
&\times&\left\{
\delta(1-\xi)\left[1+\frac{\alpha_s}{2\pi}C_F(2{\cal K}(\xi_b)-{\cal K}(2\xi_b))\right]+\frac{\alpha_s}{2\pi}C_F(1-\xi)\right\}\ ,
\end{eqnarray}
where $f_q$ represents the integrated quark distribution, the Sudakov factors resum the logarithmically enhanced contributions with the following form
\begin{eqnarray}
S^q(\zeta,{\vec b}_\perp)&=&\int_{\mu_b^2}^{\zeta^2}\frac{d{\bar\mu}^2}{{\bar\mu}^2}\left[A\ln\frac{\zeta^2}{{\bar\mu}^2}+B\right]\ ,\\
S^q_w(\zeta,\mu_L)&=&\int_{\mu_L^2}^{\zeta^2}\frac{d{\bar\mu}^2}{{\bar\mu}^2}\gamma_w \ .
\end{eqnarray}
In the above equation, we have chosen the factorization scale $\mu=\zeta$, $\xi=x_z/x'$, $\mu_b=c_0/|{\vec b}_\perp|$, 
$\mu_L=2c_0/L$. $A$ and $B$ are perturbatively calculable coefficients with 
$A=\sum_{i=1}A^{(i)}(\alpha_s/\pi)^i$ and $B=\sum_{i=1}B^{(i)}(\alpha_s/\pi)^i$, 
and the one-loop order coefficients can be read off from Eq.~(\ref{oneloop}) as $A^{(1)}=C_F/2$ and $B^{(1)}=-3C_F/4$. The additional Sudakov factor $S_w^q$ comes from the soft factor subtraction, and the anomalous dimension at one-loop is given by $\gamma_w=-C_F\frac{\alpha_s}{2\pi}$, as can be read off from the coefficient of the $\ln\frac{\zeta^2L^2}{4c_0^2}$ term in Eq.~(\ref{oneloop}). Note that we have set the scale for the Wilson line renormalization as $\mu=\zeta$ as well. In practice, it may depend on how the Wilson
lines are renormalized for the lattice computations. The hard coefficient in the second row of Eq.~(\ref{tmdaspdf}) contains all remaining one-loop contributions in Eq.~(\ref{oneloop}).

In order to carry out the matching to the usual TMDs, we compute the TMD quark distribution in the standard 
scheme~\cite{Collins,Catani:2000vq,Catani:2013tia,Prokudin:2015ysa} as well, 
\begin{eqnarray}
{q}_{TMD}(x_z,{\vec b}_\perp;\zeta^2)&=& e^{-{ {S}^q(\zeta,{\vec b}_\perp)}}\int\frac{dx'}{x'}f_q(x',\mu_b)\left\{
\delta(1-\xi)\left[1+{\cal O}(\alpha_s^2)\right]+\frac{\alpha_s}{2\pi}C_F(1-\xi)\right\}\ ,\nonumber\\
\end{eqnarray}
where $\zeta^2$ represents the hard momentum scale for the TMDs
extracted from the experiments, for example, the invariant mass of lepton
pair in the Drell-Yan lepton pair production process. We can also define
the above standard TMD as that in the Collins 2011 scheme~\cite{Collins}. 
We would like to emphasize that the Sudakov factor is the same as above. Notice that in the
standard TMD scheme (or Collins 2011 scheme), 
the hard coefficient vanishes at one-loop order. Comparing the above
two equations, we can read out the matching between the
Q-TMD quark distribution and the standard TMD quark distribution
as
\begin{eqnarray}
{q}_{QTMD}(x_z,{\vec b}_\perp;\zeta^2)&=&e^{-{ {S}_w^q(\zeta,\mu_L)}}{q}_{TMD}(x_z,{\vec b}_\perp;\zeta^2)\left[1+\frac{\alpha_s}{2\pi}C_F(2{\cal K}(\xi_b)-{\cal K}(2\xi_b))\right] \ . \label{final}
\end{eqnarray}
The above equation indicates that the Q-TMD computed on the lattice can be interpreted 
as the TMD for phenomenological applications.

\section{Discussions and Summary}

Our final result as shown in Eq.~(\ref{final}) has a number of 
interesting features. First, because the gauge links in the unsubtracted
and subtracted TMD contain Wilson line renormalization, we have
additional scale evolution expressed in term of $e^{-{ {S}_w^q(\zeta,\mu_L)}}$.
If different renormalization is chosen, we will have a different factor. For
example, for the cutoff scheme in the lattice calculation, we will have different 
factor. In practical calculations, we may not need to perform resummation
for this term at all.

In the matching coefficient, we have a  functional dependence on
$\xi_b$. Its contribution depends on the relative size between $L$ and ${\vec b}_\perp$.
In the non-perturbative region with ${\vec b}_\perp\gg L$, this is a power correction and
can be safely ignored. On the other hand, in perturbative region of ${\vec b}_\perp\ll L$, it could lead to a large
logarithm. This, however, will be dominated over by the Sudakov logs of $e^{-S^q}$.
We do no need worry too much on its contribution.
Of course, in the non-perturbative region of $L\gg {\vec b}_\perp\sim \Lambda_{\rm QCD}$,
this term may become important and needs to be carefully handled. If we can vary 
the gauge link length $L$ in such way, we may be able to avoid this region. Therefore,
this additional term does not cause any problem. 

To illustrate the above point, we have investigated the behavior of the term $2{\cal K}(\xi_b)-{\cal K}(2\xi_b)$ by plotting itas a function of ${\vec b}_\perp$ for different choices of $L$, which implies that
an optimal choice of $L$ would be around $2\sim 3/P_z$ for a reasonable range of
$P_z$.

To summarize, we have laid out the basic procedure to 
compute the TMDs from lattice QCD using LaMET or Q-TMDs. 
We have shown that the finite length of gauge links
plays a crucial role to regulate the so-called pinch pole 
singularities associated with space-like gauge links 
in the Q-TMDs. Additional soft factor subtraction improves
the theoretical convergence, especially that it cancels out
the linear divergence completely. This paves the way to correctly 
interpret the numerical results in lattice calculations of the 
TMDs. We have also shown that the energy evolution equation
for the Q-TMDs comes from the large momentum 
of the hadron $P_z$. At one-loop order, a double logarithm depending
on $P_z$ is found in the Q-TMD calculations. 
The relevant evolution equation and resummation can be
performed following the TMD formalism. 
In particular, our results show that the energy
evolution does not depend on the gauge link length $L$.

Our results may provide a justification of the technique
set up in previous attempts to calculate the TMDs on 
the lattice~\cite{Musch:2010ka}.
However, we would like to emphasize that the Q-TMD
depends on longitudinal momentum fraction $x$. Integral
over $x$ may induce difficulties to interpret the 
results from lattice calculations.

Further developments shall follow along the direction
outlined in this paper. In particular, we would like to apply our method to a realistic
calculation of the Q-TMDs on the lattice. This will be considered 
in future work. Extensions to the Wigner distributions and other 
nucleon tomography observables are desirable to follow up as well.

\vspace{2em}

We thank Markus Ebert and Iain Stewart for interesting conversations related
to the subject of this paper.
This material is based upon work partially supported by the LDRD program of 
Lawrence Berkeley National Laboratory, the U.S. Department of Energy, 
Office of Science, Office of Nuclear Physics, under contract number 
DE-AC02-05CH11231, and within the framework of the TMD Topical Collaboration,
and a grant from National Science Foundation of China (X.J.). 
JHZ is supported by the SFB/TRR-55 grant ``Hadron Physics from Lattice QCD'' and 
by a grant from National Science Foundation of China (No.~11405104). YZ is also supported by the the U.S. Department of 
Energy, Office of Science, Office of Nuclear Physics, from DE-SC0011090.

\end{document}